\documentclass[12pt]{article}

\begin{document}

\newcommand{\Mvariable}[1]{{#1}}
\newcommand{\overbar}[1]{{\bar{#1}}}
\newcommand{\imag}{i}


\newcommand{\bb}{\begin{equation}}
\newcommand{\ee}{\end{equation}}
\newcommand{\bbb}{\begin{eqnarray}}
\newcommand{\eee}{\end{eqnarray}}
\newcommand{\diag}{\mbox{diag }}
\newcommand{\Str}{\mbox{STr }}
\newcommand{\Tr}{\mbox{Tr }}
\newcommand{\Det}{\mbox{Det }}
\newcommand{\C}[2]{{\lk [{#1},{#2}\re ]}}
\newcommand{\AC}[2]{{\lk \{{#1},{#2}\re \}}}
\newcommand{\kk}{\hspace{.5em}}
\newcommand{\vc}[1]{\mbox{$\vec{{\bf #1}}$}}
\newcommand{\mc}[1]{\mathcal{#1}}
\newcommand{\del}{\partial}
\newcommand{\lk}{\left}
\newcommand{\ave}[1]{\mbox{$\langle{#1}\rangle$}}
\newcommand{\re}{\right}
\newcommand{\pd}[1]{\frac{\del}{\del #1}}
\newcommand{\pdd}[2]{\frac{\del^2}{\del #1 \del #2}}
\newcommand{\Dd}[1]{\frac{d}{d #1}}
\newcommand{\sech}{\mbox{sech}}
\newcommand{\pref}[1]{(\ref{#1})}

\newcommand
{\sect}[1]{\vspace{20pt}{\LARGE}\noindent
{\bf #1:}}
\newcommand
{\subsect}[1]{\vspace{20pt}\hspace*{10pt}{\Large{$\bullet$}}\mbox{ }
{\bf #1}}
\newcommand
{\subsubsect}[1]{\hspace*{20pt}{\large{$\bullet$}}\mbox{ }
{\bf #1}}

\def\ie{{\it i.e.}}
\def\eg{{\it e.g.}}
\def\cf{{\it c.f.}}
\def\etal{{\it et.al.}}
\def\etc{{\it etc.}}

\def\e{{\mbox{{\bf e}}}}
\def\AA{{\cal A}}
\def\BB{{\cal B}}
\def\CC{{\cal C}}
\def\DD{{\cal D}}
\def\EE{{\cal E}}
\def\FF{{\cal F}}
\def\GG{{\cal G}}
\def\HH{{\cal H}}
\def\II{{\cal I}}
\def\JJ{{\cal J}}
\def\KK{{\cal K}}
\def\LL{{\cal L}}
\def\MM{{\cal M}}
\def\NN{{\cal N}}
\def\OO{{\cal O}}
\def\PP{{\cal P}}
\def\QQ{{\cal Q}}
\def\RR{{\cal R}}
\def\SS{{\cal S}}
\def\TT{{\cal T}}
\def\UU{{\cal U}}
\def\VV{{\cal V}}
\def\WW{{\cal W}}
\def\XX{{\cal X}}
\def\YY{{\cal Y}}
\def\ZZ{{\cal Z}}

\def\sinh{{\rm sinh}}
\def\cosh{{\rm cosh}}
\def\tanh{{\rm tanh}}
\def\sgn{{\rm sgn}}
\def\det{{\rm det}}
\def\trace{{\rm Tr}}
\def\exp{{\rm exp}}
\def\sh{{\rm sh}}
\def\ch{{\rm ch}}

\def\ell{{\it l}}
\def\str{{\it str}}
\def\lp{\ell_{{\rm pl}}}
\def\blp{\overline{\ell}_{{\rm pl}}}
\def\ls{\ell_{{\str}}}
\def\bls{{\bar\ell}_{{\str}}}
\def\bM{{\overline{\rm M}}}
\def\gs{g_\str}
\def\gym{{g_{Y}}}
\def\geff{g_{\rm eff}}
\def\eff{{\rm eff}}
\def\r11{R_{11}}
\def\kel{{2\kappa_{11}^2}}
\def\kten{{2\kappa_{10}^2}}
\def\lpten{{\lp^{(10)}}}
\def\alp{{\alpha '}}
\def\alpe{{{\alpha_e}}}
\def\le{{{l}_e}}
\def\aleff{{\alp_{eff}}}
\def\sqaleff{{\alp_{eff}^2}}
\def\tgs{{\tilde{g}_s}}
\def\talp{{{\tilde{\alpha}}'}}
\def\tlp{{\tilde{\ell}_{{\rm pl}}}}
\def\tr11{{\tilde{R}_{11}}}
\def\wtilde{\widetilde}
\def\what{\widehat}
\def\hlp{{\hat{\ell}_{{\rm pl}}}}
\def\hr11{{\hat{R}_{11}}}
\def\hf{{\textstyle\frac12}}
\def\coeff#1#2{{\textstyle{#1\over#2}}}
\def\CY{Calabi-Yau}
\def\lessapprox{\;{\buildrel{<}\over{\scriptstyle\sim}}\;}
\def\greaterapprox{\;{\buildrel{>}\over{\scriptstyle\sim}}\;}
\def\inbar{\,\vrule height1.5ex width.4pt depth0pt}
\def\IC{\relax\hbox{$\inbar\kern-.3em{\rm C}$}}
\def\IR{\relax{\rm I\kern-.18em R}}
\def\IP{\relax{\rm I\kern-.18em P}}
\def\Z{{\bf Z}}
\def\R{{\bf R}}
\def\One{{1\hskip -3pt {\rm l}}}
\def\sst{\scriptscriptstyle}
\def\osc{{\rm\sst osc}}
\def\lam{\lambda}
\def\lc{{\sst LC}}
\def\pr{{\sst \rm pr}}
\def\cl{{\sst \rm cl}}
\def\D{{\sst D}}
\def\bh{{\sst BH}}
\def\vev#1{\langle#1\rangle}

\begin{titlepage}
\rightline{}

\rightline{hep-th/yymmddd}

\vskip 2cm
\begin{center}
\Large{{\bf Closed strings\\ in Ramond-Ramond backgrounds
}}
\end{center}

\vskip 2cm
\begin{center}
Vatche Sahakian\footnote{\texttt{sahakian@hmc.edu}}
\end{center}
\vskip 12pt
\centerline{\sl Keck Laboratory}
\centerline{\sl Harvey Mudd}
\centerline{\sl Claremont, CA 91711, USA}

\vskip 2cm

\begin{abstract}
We write the IIB Green-Schwarz action
in certain general classes of
curved backgrounds threaded with Ramond-Ramond fluxes. 
The fixing of the kappa symmetry in the light-cone
gauge and the use of supergravity Bianchi identities
simplify the task. We find an expression 
that truncates to quartic order in the spacetime spinors and relays
interesting information about the vacuum structure of the worldsheet theory.
The results are particularly useful in exploring integrable string dynamics
in the context of the holographic duality.

\end{abstract}

\end{titlepage}
\newpage
\setcounter{page}{1}

\section{Introduction and Results}
\label{intro}

In many realizations of 
the holographic duality~\cite{MALDA1,KLEB,WITHOLO}, 
where a perturbative
string theory is found dual to strongly coupled dynamics in a field theory
or in 
another string theory, the closed strings on the weakly coupled
side of the duality are immersed in background Ramond-Ramond (RR) fluxes.
Knowledge of the couplings of the string  worldsheet degrees of freedom to such fluxes is then an important 
ingredient to the task of exploring the  underpinnings of the duality.

There are three main
approaches in writing down an action of closed superstrings in
an arbitrary background. In the RNS formalism, powerful 
computational techniques
are available, yet the vertex operators sourced by RR fields involve
spin fields. A second approach is the Green-Schwarz (GS) formalism with spacetime
supersymmetry, generally leading
to an action that is particularly useful in unraveling the
semi-classical dynamics of the sigma model. On the down side, manifest
Lorentz symmetry is lost with the fixing of the light-cone gauge; and,
at one loop level for example, the lost symmetry results in
serious complications. The third approach was developed 
recently~\cite{BVW}
and involves a hybrid picture. In this strategy, part of the spacetime
symmetries remain manifest yet couplings to the RR fields take 
relatively simple forms.
The cost is the introduction of several auxiliary fields, 
and certain assumptions on the form of the background.

In this work, we focus on the second GS formalism with spacetime supersymmetry
and on determining the component form of this action. Our interest is to
eventually study, semi-classically, closed string dynamics in general
backgrounds that arise readily in the study of D-branes. Other recent
approaches involve specializing to backgrounds with a large amount of symmetry in writing the corresponding worldsheet theory; for example, AdS spaces have been of particular interest (e.g. \cite{RAHMRAJ}-\cite{PESANDO}). We would then like to 
extend the scope of this program by considering generic D-brane configurations with much less symmetry. 

Most of the difficulties involved in writing down the string action
in general form are due to the fact that superspace
for supergravity, while still being a natural
setting for the theory, can be
considerably complicated~\cite{SUSYBOOK}: 
a large amount of superfluous symmetries need to
be fixed and computations are often prohibitively lengthy.

The task is significantly simplified by the use of the method
of normal coordinate
expansion~\cite{MUKHINLSM,GRISARU} 
in superspace. This was developed for the Heterotic
string in~\cite{ATICKDHAR}, and, along with the use of computers for analytical
manipulations, makes determining the type IIA and IIB 
sigma models straightforward as well. The additional complications
that arise in these cases - and that are absent in the Heterotic string case - 
are due entirely to the presence of the RR fields. 

In this paper, we concentrate on the IIB theory. In~\cite{CVETIC}, 
part of this action, to quadratic order in the spinors, was derived
starting from the supermembrane action and using T-duality\footnote{see also~\cite{PEDRO}
for a derivation of the action to quadratic order using T-duality.}. 
In this work, starting from
IIB superspace directly and using the method of normal coordinate expansion, we compute the full form of the IIB worldsheet action in the light-cone gauge relevant to most backgrounds of interest. 
In the subsequent subsection, we present all the results of this work in a self-contained format. The details of deriving
the action are then left for the rest of the paper and need not be consulted.

\subsection{{The  results}}

The class of background fields we focus on is inspired by~\cite{ATICKDHAR} 
and by the need to apply our results to settings that arise in
the context of the holographic duality. In particular, 
fields generated by electric and magnetic
D-branes of various configurations share certain general features of interest.
We list all the conditions we impose on the background fields so that
our form of the IIB action is valid:
\begin{itemize}
\item The supergravity fermions are to vanish. In particular, 
the gaugino and gravitino have no condensates.

\item We choose a certain space direction that, along with the time
coordinate, we will associate with the light-cone gauge fixing later.
We refer to the other eight spatial directions as being transverse. With this convention, we demand that all background fields depend only on the transverse coordinates.

\item Tensor fields can have indices in the transverse directions;
and in the two light-cone directions only if the light-cone coordinates
appear in pairs; e.g. a 3-form field strength $F_{pqr}$ can have nonzero 
components $F_{abc}$ or $F_{-+a}$ where $a$, $b$, and $c$ are
transverse directions; and $+$ and $-$ are light-cone directions. But components
such as $F_{-ab}$ are zero.
\end{itemize}

For example, if we were to consider a background 
consisting of a number of static D$p$ branes,
we choose the light-cone directions {\em parallel} to the worldvolume
of the branes. All conditions listed above are then satisfied. 

Under these assumptions, and once the $\kappa$ symmetry is fixed,
the IIB action takes the form
\bb\label{genform}
\SS=\SS^{(0)}+\SS^{(2)}+\SS^{(4)}\ ,
\ee
where the superscripts denote the number of spinors in each expression.
Hence, the action truncates to quartic order in the fermions.

The first term is the standard bosonic part\footnote{
The signature of the metric we use is $(+----\cdots)$. See Appendix
A for the details.}
\bb\label{I0}
\SS^{(0)}=\int d^2\sigma\lk[ 
\frac{1}{2}\sqrt{-h}\,h^{ij} V_i^a V_{a\, j}+2\,\sqrt{-h}\, h^{ij}\, V_i^+\, V_j^-
+\frac{1}{2} \varepsilon^{ij}\,V_i^a\,V_j^b\,b^{(1)}_{ab}
+2\varepsilon^{ij} V^+_i\, V^-_j\, b^{(1)\,-+}
\re]\ .
\ee
In this expression, and throughout, 
the $'+'$ and $'-'$ flat tangent
space labels refer to the light-cone directions as in
$x^\pm\equiv (x^0\pm x^1)/2$, with $x^0$ and $x^1$ being respectively the
time and some chosen space direction defining the light-cone.
We then denote eight flat transverse
tangent space indices by $a,b,...$. 
All tensors are written with their
indices in the tangent space by using the vielbein; \ie\ for the NSNS B-field, we have
$b^{(1)}_{ab}=e_a^m e_b^n b^{(1)}_{mn}$.  
We define
\bb
V_i^a\equiv \del_i x^m e_m^a\ \ \ ,\ \ \ V_i^\pm\equiv \del_i x^m e_m^\pm\ .
\ee
Curved spacetime indices are then labeled by $m,n,...$. 
Note that we write the action in 
the Einstein frame; in section 4.1 we cast part of the action into the string frame
to compare with the literature. 

Note also that while we fix the kappa symmetry, we
do not fix the light-cone gauge in $V^+_i$ and $h_{ij}$ so as to allow for
different choices. One conventional choice in flat space is
$h_{\tau\tau}=+1$, $h_{\sigma\sigma}=-1$, $h_{\tau\sigma}=0$, $V^+_\sigma=0$, and
$V^+_\tau=p^+$.

Next, we represent the two spacetime spinors by a single Weyl -- but otherwise
complex -- 16 component spinor $\theta$. The $16\times 16$ gamma matrices are
denoted by $\sigma^a$ and the conventions for the spinor representation we have adopted are summarized in Appendix A.
At quadratic order in $\theta$, the action takes the form
\bb\label{I2}
\SS^{(2)}=\int d^2\sigma\,\lk(\II_{kin}+V^{+\,i}\, V_{c}^j\,\II_{ij}^c\re)+\mbox{c.c.}\ ,
\ee
where `c.c' stands for complex conjugate. And we define separately the kinetic
piece and the piece that involves no derivatives of the fermions
\bb
\II_{kin}=-i\, \omega \sqrt{-h}\, h^{ij} V^+_i \overbar{\theta} \sigma^- D_j\theta
+i\,\omega\, \varepsilon^{ij} V^+_i\,\theta\sigma^-D_j\theta\ ;
\ee

\bbb\label{I2fields}
\II_{ij}^c&=&i\,\frac{\omega}{2}\,  \sqrt{-h}\, h_{ij}\, P^c\, \overbar{\theta}\sigma^- \theta
-i\,\frac{\omega}{2}\varepsilon_{ij}\, P_a\, \overbar{\theta}\sigma^{-c a}\overbar{\theta} \nonumber \\
&-&i \omega \sqrt{-h}\,h_{ij}\,\,F^{-+}_{\ \ \ \,a} \overbar{\theta}\sigma^{-ca}\overbar{\theta}
+i\frac{\omega}{4} \sqrt{-h}\,h_{ij}\,F_{ab}^{\ \ \ c} \overbar{\theta}\sigma^{-ab}\overbar{\theta} \nonumber \\
&+&i\omega \varepsilon_{ij} F^{-+c} \overbar{\theta}\sigma^-\theta
-i\omega \varepsilon_{ij} F^{-+}_{\ \ \ a} \overbar{\theta} \sigma^{-ca}\theta
-i \frac{\omega}{4}\,\varepsilon_{ij} F^c_{\ ab} \overbar{\theta}\sigma^{-ab} \theta\nonumber \\
&+& \frac{\omega}{4} \varepsilon_{ij} G^{-+c}_{\ \ \ \ ab} \theta \sigma^{-ab} \theta\ .
\eee
The covariant derivative is defined as
\bb\label{Dtheta}
D_j \theta\equiv \del_j \theta^\alpha 
-\frac{1}{4} \del_j x^m
\omega_{m,ab}\,\sigma^{ab}\theta \ .
\ee
The various background fields appearing in~\pref{I2fields} are:
\begin{itemize}
\item The IIB dilaton
\bb
\omega\equiv e^{\phi/2}\ .
\ee

\item The field strengths for the IIB scalars
\bb
P_a\equiv \frac{e^\phi}{2} \lk(i D_a \chi- e^{-\phi} D_a\phi\re)\ ;
\ee
\bb
Q_a\equiv\frac{\overbar{P}_a-P_a}{4\,i}=-\frac{e^\phi}{4} D_a \chi\ ,
\ee
with $\chi$ being the IIB axion.

\item The 3-form field strength 
\bb
F_{abc}\equiv
\frac{e^{\phi/2}}{2}
(1+e^{-\phi}+i\chi)\FF_{abc}
+\frac{e^{\phi/2}}{2}(-1+e^{-\phi}+i\chi)\bar{\FF}_{abc}
\ ;
\ee
with
\bb
\FF_{abc}\equiv \frac{h^{(1)}_{abc}}{2}+i\frac{h^{(2)}_{abc}}{2}\ ,
\ee
where $h^{(1)}$ and $h^{(2)}$ are, respectively, the field
strengths associated with fundamental string and D-string charges.

\item And the five-form self-dual field strength $G_{abcde}$.
\end{itemize}

At quartic order in the spinors, the action involves many more terms. We may 
organize these in eight different parts:
\bb\label{I4}
\SS^{(4)}=\int d^2\sigma\, \sqrt{-h}\, h^{ij} V^+_i V^+_j \lk[\II_{FF}+\II_{FG}+\II_{GG}+\II_{DF}+\II_{FP}+\II_{DG}+\II_R+\II_{PP}\re]+\mbox{c.c.}
\ee
according to field content. Amongst these, we encounter two qualitatively different
types of terms: ones involving the form $\theta\bar{\theta} \theta \bar{\theta}$; and ones involving the structure $\theta\theta\theta\bar{\theta}$ (or its complex conjugate). The fermions count a $U(1)$ charge which is part of the symmetry of the supergravity theory.
We then get the expressions of the first type
\begin{eqnarray}
\II_{GG}&=&\frac{23\, \omega}{4608}\,{\lk( \overbar{\theta}\,{\sigma }^{-}\,\theta \re) }^2\,\ 
{{G^{-\,+\,{a}\,{b}\,{c}}\ G^{-\,+}_{\ \ \ \,{a}\,{b}\,{c}}}\,
     }  \nonumber \\ &+& 
  \frac{\omega}{9216}
  {\overbar{\theta}\,{\sigma }^{-\,{a}\,{b}}\,\theta }\ \,{\overbar{\theta}\,{\sigma }^{-}_{\ \ \,{a}\,{b}}\,\theta }\,\ 
  {{G^{-\,+\,{c}\,{d}\,{e}}\ G^{-\,+}_{\ \ \ \,{c}\,{d}\,{e}}}\,
     } 
     \nonumber \\ 
     &-& \frac{\omega}{384} \overbar{\theta}\,{\sigma }^{-}\,\theta\  \,
      \overbar{\theta}\,{\sigma }^{-\,{a}\,{b}}\,\theta
      \left[
  \frac{1}{12}{{G}^{-\,+\,{c}\,{d}\,{e}}\,{G}_{{c}\,{d}\,{e}\,{a}\,{b}}\,
        } + 
 {{G}^{-\,+\,{c}}_{\ \ \ \ \ \,{a}\,{d}}\,
     {G}^{-\,+\,{d}}_{\ \ \ \ \ \,{\,{c}\,{b}}} \,
        } \right]\nonumber \\  
      &+& \frac{\omega}{1536} \overbar{\theta}\,{\sigma }^{-\,{a}\,{b}}\,\theta\  \,
      \overbar{\theta}\,{\sigma }^{-\,{c}\,{d}}\,\theta 
  \left[{{G}^{-\,+\,{e}}_{\ \ \ \ \ \,{a}\,{b}}\,
     {G}^{-\,+}_{\ \ \ \,{e}\,{c}\,{d}} \,
        }  
     -  \frac{1}{24}{{G}_{{a}\,{b}\,{e}\,{f}\,{g}}\,{G}^{{e}\,{f}\,{g}}_{\ \ \ \ \,{c}\,{d}}} 
    \right] \nonumber \\ 
     &-& \frac{\omega}{256} \overbar{\theta}\,{\sigma }^{-\,{a}\,{c}}\,\theta\  \,
      \overbar{\theta}\,{\sigma }^{-\,{b}}_{\ \ \ \ \,{c}}\,\theta 
      \left[
  {{G}^{-\,+}_{\ \ \ \ \,{a}\,{d}\,{e}}\,
     {G}^{-\,+\,{d}\,{e}}_{\ \ \ \ \ \ \ \,{b}}}-
\frac{1}{72}{{G}_{{a}\,{d}\,{e}\,{f}\,{g}}\,{G}^{{d}\,{e}\,{f}\,{g}}_{\ \ \ \ \ \ \,{b}} \,
       }\right]
\end{eqnarray}
\begin{eqnarray}
\II_{FF}&=&
\frac{13\,\omega}{24}\,{\lk( \overbar{\theta}\,{\sigma }^{-}\,\theta \re) }^2\,\ 
{{F}^{-\,+\,{a}}\,{\overbar{F}}^{-\,+}_{\ \ \ \ \,{a}}} + 
  \frac{25\,\omega}{768}\,{ \overbar{\theta}\,{\sigma }^{-\,{a}\,{b}}\,\theta }\ { \overbar{\theta}\,{\sigma }^{-}_{\ \,{a}\,{b}}\,\theta  }\,\ 
  {{F}^{-\,+\,{c}}\,{\overbar{F}}^{-\,+}_{\ \ \ \ \,{c}} \,
     } \nonumber \\ &+& \frac{\omega}{256}\,
     \overbar{\theta}\,{\sigma }^{-}\,\theta\  \,
      \overbar{\theta}\,{\sigma }^{-\,{a}\,{b}}\,\theta  
      \left[
  93\,{{F}^{-\,+}_{\ \ \ \ \,{a}}\,{\overbar{F}}^{-\,+}_{\ \ \ \ \,{b}} \,
      } - 
  \frac{43}{2}\,{{F}^{-\,+\,{c}}\,{\overbar{F}}_{{c}\,{a}\,{b}} \,
      } - 
  \frac{17}{24}\,{{F}_{{a}\,{c}\,{d}}\,{\overbar{F}}^{{c}\,{d}}_{\ \ \ \,{b}} \,
      }\right] \nonumber \\ &-& \frac{\omega}{96}\, 
  \overbar{\theta}\,{\sigma }^{-\,{a}\,{c}}\,\theta\  \,
      \overbar{\theta}\,{\sigma }^{-\,{b}}_{\ \ \ \ \,{c}}\,\theta  
      \left[17\,{{F}^{-\,+}_{\ \ \ \ \,{a}}\,{\overbar{F}}^{-\,+}_{\ \ \ \ \,{b}} \,
      }+ 
  \frac{5}{8}\,{{F}^{-\,+\,{d}}\,{\overbar{F}}_{{d}\,{a}\,{b}} \,
      } - 
  \frac{7}{16}\,{{F}_{{a}\,{d}\,{e}}\,{\overbar{F}}^{{d}\,{e}}_{\ \ \ \,{b}} \,
      }\right]
      \nonumber \\ &+& \frac{\omega}{1536}\, 
  \overbar{\theta}\,{\sigma }^{-\,{a}\,{b}}\,\theta\  \,
      \overbar{\theta}\,{\sigma }^{-\,{c}\,{d}}\,\theta  
      \left[ 23\,{{F}^{-\,+}_{\ \ \ \ \,{d}}\,{\overbar{F}}_{{c}\,{a}\,{b}} \,
      }- 
  7\,{{F}^{-\,+}_{\ \ \ \ \,{a}}\,{\overbar{F}}_{{b}\,{c}\,{d}} \,
      } \right. \nonumber \\ &-& \left. 
  \frac{1}{2}{{F}_{{a}\,{c}\,{e}}\,{\overbar{F}}^{{e}}_{\ \,{b}\,{d}} \,
      }- 
\frac{13}{4}\,{{F}_{{a}\,{b}\,{e}}\,{\overbar{F}}^{{e}}_{\ \,{c}\,{d}} \,
      } \right]
  \end{eqnarray}
\begin{eqnarray}
\II_{DG}&=&\frac{\imag }{192}\,\omega \,{D_{{c}}}\,G^{-\,+\,{c}}_{\ \ \ \ \ \,{a}\,{b}}  \,
    \overbar{\theta}\,{\sigma }^{-}\,\theta  \,
    \overbar{\theta}\,{\sigma }^{-\,{a}\,{b}}\,\theta 
\end{eqnarray}
\begin{eqnarray}
\II_{PP}&=&\frac{15\,\omega}{256}\,\overbar{\theta}\,{\sigma }^{-}\,\theta  \,\ 
      \overbar{\theta}\,{\sigma }^{-\,{a}\,{b}}\,\theta\,\ 
      {{P}_{{a}}\,{\overbar{P}}_{{b}}} - 
  \frac{\omega}{48}\,\overbar{\theta}\,{\sigma }^{-\,{a}\,{c}}\,\theta  \,\ 
      \overbar{\theta}\,{\sigma }^{-\,{b}}_{\ \ \ \ \,{c}}\,\theta\,\ 
      {{P}_{{a}}\,{\overbar{P}}_{{b}}}
\end{eqnarray}
\begin{eqnarray}
\II_{R}&=&-\frac{5\,\omega}{32}\,\ 
{\lk( \overbar{\theta}\,{\sigma }^{-}\,\theta \re) }^2\, {R^{-\,+\,-\,+} \,
     }  + 
  \frac{\omega}{192}\,\ 
  {\overbar{\theta}\,{\sigma }^{-\,{a}\,{b}}\,\theta}\ {\overbar{\theta}\,{\sigma }^{-}_{\ \,{a}\,{b}}\,\theta}\,\ 
  {R^{-\,+\,-\,+}}  \nonumber \\ &-& \frac{\omega}{96}\, 
     \overbar{\theta}\,{\sigma }^{-\,{a}\,{c}}\,\theta  \,\ 
      \overbar{\theta}\,{\sigma }^{-\,{b}}_{\ \ \ \,{c}}\,\theta  
      \left[
     {R^{-\,+}_{\ \ \ \ \,{a}\,{b}} \,
      } +
  \frac{1}{2}{R_{{a}\,{d}\,{b}}^{\ \ \ \ \,{d}}  
     }\right] \nonumber \\ &+& \frac{\omega}{384}\, 
      \overbar{\theta}\,{\sigma }^{-\,{a}\,{b}}\,\theta  \,\ 
      \overbar{\theta}\,{\sigma }^{-\,{c}\,{d}}\,\theta  
       \left[
  {R_{{a}\,{c}\,{b}\,{d}} 
     }+
\frac{1}{2}{R_{{a}\,{b}\,{c}\,{d}}}\right] 
\end{eqnarray}
$R_{abcd}$ being the Riemann tensor in the Einstein frame.

The expressions of the second type are
\begin{eqnarray}\label{new1}
\II_{FG}&=& i \frac{\omega }{48}\,\theta\,{\sigma }^{-\,{a}\,{b}}\,\theta\,\ \,\overbar{\theta}\,{\sigma }^{-}\,\theta 
\,{\overbar{F}}^{-\,+\,{e}}\,G^{-\,+}_{\ \ \ \,{a}\,{b}\,{e}}\,
     \,
       \nonumber \\
&+& i \frac{\omega}{32}\,  \theta\,{\sigma }^{-\,{a}\,{b}}\,\theta\,\ \,
    \overbar{\theta}\,{\sigma }^{-\,{c}\,{d}}\,\theta
    \left[
    \frac{1 }{3}\,{\overbar{F}}_{{e}\,{b}\,{d}}\,G^{-\,+\,{e}}_{\ \ \ \ \ \,{a}\,{c}}\, \,
       -
  \frac{1 }{3}\,{\overbar{F}}^{-\,+}_{\ \ \ \ \,{d}}\,G^{-\,+}_{\ \ \ \,{a}\,{b}\,{c}}\,
       - 
  \frac{5}{12}\,{\overbar{F}}_{{e}\,{c}\,{d}}\,G^{-\,+\,{e}}_{\ \ \ \ \ \,{a}\,{b}}\, \right.   \nonumber \\ &-&\left.
  \frac{1}{12}\,{\overbar{F}}_{{e}\,{f}\,{d}}\,G^{{e}\,{f}}_{\ \ \,{a}\,{b}\,{c}}\,     + 
  \frac{1}{12}\,{\overbar{F}}_{{e}\,{a}\,{b}}\,G^{-\,+\,{e}}_{\ \ \ \ \ \,{c}\,{d}}\,    -
  \,{\overbar{F}}^{-\,+}_{\ \ \ \ \,{b}}\,G^{-\,+}_{\ \ \ \,{a}\,{c}\,{d}}\,
       + 
  \frac{1}{3}\,{\overbar{F}}^{-\,+}_{\ \ \ \ \,{e}}\,G^{{e}}_{\ \,{a}\,{b}\,{c}\,{d}}\,    \right] 
    \nonumber \\ 
    &+& i \frac{\omega}{48}\, \theta\,{\sigma }^{-\,{a}\,{c}}\,\theta\,\ \,
    \overbar{\theta}\,{\sigma }^{-\,{b}}_{\ \ \ \ \,{c}}\,\theta
    \left[
  {\overbar{F}}^{-\,+}_{\ \ \ \ \,{d}}\,G^{-\,+\,{d}}_{\ \ \ \ \ \,{a}\,{b}}\,
       + \frac{1}{4}\,{\overbar{F}}_{{d}\,{e}\,{b}}\,G^{-\,+\,{d}\,{e}}_{\ \ \ \ \ \ \ \,{a}} \,
       -
  \frac{1}{4}\,{\overbar{F}}_{{d}\,{e}\,{a}}\,G^{-\,+\,{d}\,{e}}_{\ \ \ \ \ \ \ \,{b}} \,
      \right] \nonumber \\ &+& 
  i \frac{\omega }{1152}\,
  \theta\,{\sigma }^{-\,{a}\,{b}}\,\theta\,\ \,
    \overbar{\theta}\,{\sigma }^{-}_{\ \ \,{a}\,{b}}\,\theta\,
    {\overbar{F}}_{{c}\,{d}\,{e}}\,G^{-\,+\,{c}\,{d}\,{e}} \, 
\end{eqnarray}
\begin{eqnarray}\label{new2}
\II_{FP}&=&
\frac{\omega}{8}\,{{F}^{-\,+}_{\ \ \ \ \,{a}}\,{\overbar{P}}_{{b}}\ \,
      \overbar{\theta}\,{\sigma }^{-}\,\theta\  \,
      \theta\,{\sigma }^{-\,{a}\,{b}}\,\theta  } - 
  \frac{\omega}{8}\,{{F}^{-\,+}_{\ \ \ \ \,{a}}\,{\overbar{P}}_{{b}}\ \,
      \overbar{\theta}\,{\sigma }^{-\,{a}\,{c}}\,\theta\  \,
      \theta\,{\sigma }^{-\,{b}}_{\ \ \ \,{c}}\,\theta  } \nonumber \\ &+& 
  \frac{\omega}{96}\,{{F}_{{a}\,{c}\,{d}}\,{\overbar{P}}_{{b}}\ \,
      \overbar{\theta}\,{\sigma }^{-\,{c}\,{d}}\,\theta\  \,
      \theta\,{\sigma }^{-\,{a}\,{b}}\,\theta} \nonumber \\
      &-&
i\,\frac{\omega}{6} \, \theta\,{\sigma }^{-\,{a}\,{c}}\,\theta\,\ 
      \overbar{\theta}\,{\sigma }^{-\,{b}}_{\ \ \ \ \,{c}}\,\theta  \,\ 
      {{Q_{{b}}}\,{\overbar{F}}^{-\,+}_{\ \ \ \ \,{a}}  \,
        } - 
  i\,\frac{\omega}{24} \, \theta\,{\sigma }^{-\,{a}\,{b}}\,\theta\,\ 
      \overbar{\theta}\,{\sigma }^{-\,{c}\,{d}}\,\theta  \, \ 
      {{Q_{{c}}}\,{\overbar{F}}_{{a}\,{b}\,{d}}
      }
\end{eqnarray}
\begin{eqnarray}\label{new3}
\II_{DF}&=&
\frac{\omega}{12} \, \theta\,{\sigma }^{-\,{a}\,{c}}\,\theta\,\ 
      \overbar{\theta}\,{\sigma }^{-\,{b}}_{\ \ \ \ \,{c}}\,\theta  \,\ 
      {{D_{{b}}}\,{\overbar{F}}^{-\,+}_{\ \ \ \ \,{a}}  \,
        } + 
  \frac{\omega}{48} \, \theta\,{\sigma }^{-\,{a}\,{b}}\,\theta\,\ 
      \overbar{\theta}\,{\sigma }^{-\,{c}\,{d}}\,\theta  \, \ 
      {{D_{{c}}}\,{\overbar{F}}_{{a}\,{b}\,{d}}
      }
\end{eqnarray}

To make contact with
the literature, we write the equations of
motion satisfied by the background fields in the conventions we have adopted. 
Labeling flat tangent space indices that span all ten spacetime directions (including
the light-cone) by $\hat{a},\hat{b},\hat{c},\hat{d},\ldots$,
we have~\cite{HOWEWEST}
\bb
D^{\hat{a}} P_{\hat{a}}=-4\,i\, Q^{\hat{a}} P_{\hat{a}}+\frac{1}{6} F_{\hat{a}\hat{b}\hat{c}} F^{\hat{a}\hat{b}\hat{c}}\ ;
\ee
\bb
D^{\hat{c}} F_{\hat{a}\hat{b}\hat{c}}=-2\, i\,Q^{\hat{c}} F_{\hat{a}\hat{b}\hat{c}}-\overbar{F}_{\hat{a}\hat{b}\hat{c}} P^{\hat{c}}+\frac{i}{6} G_{\hat{a}\hat{b}\hat{c}\hat{d}\hat{e}} F^{\hat{c}\hat{d}\hat{e}}\ ;
\ee
\bb
D^{\hat{e}} G_{\hat{a}\hat{b}\hat{c}\hat{d}\hat{e}}=-\frac{i}{18} \varepsilon_{\hat{a}\hat{b}\hat{c}\hat{d}\hat{e}\hat{f}\hat{g}\hat{h}\hat{i}\hat{j}} F^{\hat{e}\hat{f}\hat{g}} \overbar{F}^{\hat{h}\hat{i}\hat{j}}
\ee
\bb
R_{\hat{a}\hat{b}}=-2 \overbar{P}_{\lk(\hat{a}\re.} P_{\hat{b})}-\overbar{F}_{(\hat{a}}^{\ \ \hat{c}\hat{d}} F_{\hat{b})}^{\ \ \hat{c}\hat{d}}
+\frac{1}{12} \eta_{\hat{a}\hat{b}} \overbar{F}_{\hat{c}\hat{d}\hat{e}} F^{\hat{c}\hat{d}\hat{e}}
-\frac{1}{96} G_{\hat{a}}^{\ \hat{c}\hat{d}\hat{e}\hat{f}} G_{\hat{b}\hat{c}\hat{d}\hat{e}\hat{f}}\ ;
\ee
The five-form field strength $G_{\hat{a}\hat{b}\hat{c}\hat{d}\hat{e}}$ is self-dual in this scheme.
Our conventions conform, for example,
to those in~\cite{SL2Z} with the identifications $\chi+i\,e^{-\phi}\rightarrow\lambda$, 
$h^{(1)}\rightarrow H^{(1)}$,
$h^{(2)}\rightarrow H^{(2)}$.

In the rest of the paper, we
describe how to arrive at the expressions presented. 
The current section was organized
such that the details of these derivations
are not needed to make use of the results. Section 2 summarizes the basic
superspace formalism we use. Section 3 outlines the strategy and techniques that
simplify the computations. Section 4 presents yet more details; in particular, in section 4.1, 
we write the action in the string frame to quadratic order in the spinors, compare with the literature,
and confirm that our results are consistent with other recent attempts at determining
this action (see however minor note in section 4.1 with regards to the U(1) charge). Section 5 includes concluding remarks about future directions. And Appendix A collects some of the conventions we use throughout the paper.

\vspace{0.3in}
\begin{center}
NOTE ADDED
\end{center}

\vspace{0.1in}
The original version of this paper~\cite{VVSOLD} presented the action to quartic order
but without the terms involving the spinor structure $\theta\theta\theta\overbar{\theta}$. 
In that version, it was erroneously argued - as pointed out by~\cite{OTHER} - that these terms would vanish. In this work, this argument
has been corrected and the additional terms are now presented in equations~\pref{new1}-\pref{new3}.
Furthermore, the entire computation has been reworked and organized
so as to make the calculation of all the terms sensitive to the same
potential human errors. Since we are now able to check against the literature
for the terms quadratic in the fermions, this organization of the computation provides
an indirect check of the quartic terms as well. The entire computational
scheme has been coded on Mathematica 5.0. 

\section{Preliminaries}

\subsection{IIB supergravity in superspace}

The fields of
IIB supergravity are
\bb\label{fields}
\lk\{
{e}^{\hat{a}}_m,\  \tau=e^{-\phi}+i \chi,\  b^{(1)}_{mn} + i\  b^{(2)}_{mn},\ 
b_{mnrs};\  \psi_m,\ \lambda
\re\}\ ;
\ee
these are respectively
the vielbein, a complex scalar comprised of the
dilaton and the axion, two two-form gauge fields, a four-form real gauge field,
a complex left-handed gravitino, and a complex right-handed spinor. 
The gauge fields have the associated field strengths defined as
\bb
h^{(1)}=db^{(1)}\ ,\ \ \ 
h^{(2)}=db^{(2)}\ ,\ \ \ 
g=db\ .
\ee

An elaborate superspace formalism can be developed for this theory.
It involves the standard supergravity superfields~\cite{HOWEWEST}
\bb\label{superfields1}
\lk(E^A_M,\Omega_{MA}^{B}\re)\rightarrow
\lk(T^A_{BC},R_{ABC}^{D}\re)\ .
\ee
In addition, one needs five other tensor superfields
\bb\label{superfields2}
\lk\{P_A,Q_A,{\hat{\FF}}_{ABC},\GG_{ABCDE},\Lambda_A\re\}
\ee
Throughout, we accord
to the standard convention of denoting tangent space
superspace indices by capital
letters from the beginning of the alphabet. 
In this setting of $\NN=2$ chiral supersymmetry, 
an index such as $A$ represents a tangent space 
vector index $\hat{a}$ spanning all ten dimensions, and two spinor
indices $\alpha$ and $\bar{\alpha}$.
Hence, superspace is
parameterized by coordinates
\bb
z^A\in\lk\{x^{\hat{a}},\theta^\alpha,\theta^{\bar{\alpha}}\re\}\ .
\ee
Here, $\theta^\alpha$ and $\theta^{\bar{\alpha}}\equiv {\bar{\theta}}^\alpha$
have same chirality and are related to each other by complex conjugation. 
In this manner, 
unbarred and barred Greek letters from the beginning of the alphabet
will be used to denote spinor indices. 
More details about the conventions we adopt can be found in Appendix A.

The two superfields $P_A$ and $Q_A$ are the field strengths of a matrix
of scalar superfields
\bb
\VV=\lk(
\begin{array}{cc}
u & v \\
\overline{v} & \overline{u}
\end{array}
\re)
\ ,
\ee
with
\bb
u\overline{u}-v\overline{v}=1\ .
\ee
This matrix describes the group $SU(1,1)\sim SL(2,R)$, which
later gets identified with the S-duality group of the IIB theory.
The scalars parameterize the coset space $SU(1,1)/U(1)$, with the additional
$U(1)$ being a space-time dependent symmetry with an associated gauge field.
We then define
\bb
\VV^{-1} d\VV\equiv \lk(
\begin{array}{cc}
2 i Q & P \\
\overline{P} & -2 i Q
\end{array}
\re)\ ,
\ee
with
\bb
Q=\overline{Q}
\ee
being the $U(1)$ gauge field mentioned above. All fields in the theory
carry accordingly various charge assignments under this $U(1)$. This is
a powerful symmetry that can be used to severely restrict the superspace
formalism.
We also introduce the superfield strength $\hat{F}$
\bb
\lk( \bar{{\hat{\FF}}}, \hat{\FF}\re)=
\lk(\bar{{\hat{F}}},\hat{F}\re)\VV^{-1}\ ,
\ee
which transforms under the $SU(1,1)$ as a singlet.

All these fields are associated with a myriad of Bianchi identities.
As is typical in supergravity theories, there is an immense amount of
superfluous symmetries in the superspace formalism. Some of these
can be fixed conventionally; and using the Bianchi identities, relations
can be derived between the various other components. 
We will be very brief in reviewing this formalism, as 
our focus will be the string sigma model. Instead of reproducing
the full set of equations that determine the IIB theory, we present 
only those statements that are of direct relevance to the
worldsheet theory.
Throughout this work,
we accord closely to the conventions and notation
of \cite{HOWEWEST}; the reader may refer
at any point to~\cite{HOWEWEST} to complement his/her reading.

From the point of view of the IIB string sigma model, the following
combination of the scalars turns out to play an important role
\bb
\omega=u-\overline{v}\ .
\ee
Requiring $\kappa$ symmetry on the worldsheet leads to the condition
\bb\label{gauge}
\omega=\bar{\omega}\ .
\ee
This is a choice that is unconventional from the point of view
of the supergravity formalism, but is natural from the perspective
of the string sigma model.

We parameterize the scalar superfields as~\cite{RADAK,BELLUCI}
\bb
u=\frac{1+\bar{W}}{\sqrt{2 (W+\bar{W})}} e^{-2 i \theta}\ ,
\ee
\bb
v=-\frac{1-W}{\sqrt{2 (W+\bar{W})}} e^{2 i \theta}\ ,
\ee
with the three variables $W$, $\bar{W}$ and $\theta$ parameterizing the
$SU(1,1)$. The gauge choice~\pref{gauge} then corresponds to
\bb
\theta=0\ ,
\ee
This leads to
\bb
\omega=\sqrt{\frac{2}{W+\bar{W}}}\ .
\ee
And
\bb
Q_A=\frac{\bar{P}_A-P_A}{4 i}\ .
\ee
Finally, the field strengths are given in terms of $W$ by
\bb
P=\frac{dW}{W+\bar{W}}\ ,\ \ \ 
Q=\frac{i}{4} \frac{d(W-\bar{W})}{W+\bar{W}}\ .
\ee

To make contact with the IIB theory's field content~\pref{fields},
we need to specify the map between the superfields~\pref{superfields1} and
~\pref{superfields2} and the
physical fields. Each superfield involves an expansion in the fermionic
superspace coordinates $\theta$. At zeroth order in this expansion, we 
have
\bb
W|_0=\tau=e^{-\phi}+i \chi\ ,
\ee
Similarly, the zeroth components of the $\Lambda$ superfield is
\bb
\Lambda_\alpha|_0=\lambda_\alpha\ .
\ee
In the Wess-Zumino gauge, the supervielbein's zeroth component is
\bb
E_{m}^\alpha|_0=\psi_m^\alpha\ .
\ee
At this point, we can simplify the discussion significantly by 
choosing to set all background fermionic fields to zero
\bb
\lambda_\alpha\rightarrow 0\ ,\ \ \ 
\psi_m^\alpha\rightarrow 0\ .
\ee
This identifies the class of backgrounds which is of most interest
to us and  that arises most frequently in the literature.
Given this, the zeroth components of the other fields are
\bb
{\hat{\FF}}_{\hat{a}\hat{b}\hat{c}}|_0=\FF_{\hat{a}\hat{b}\hat{c}}\equiv
\frac{h^{(1)}_{\hat{a}\hat{b}\hat{c}}}{2}
+i \frac{h^{(2)}_{\hat{a}\hat{b}\hat{c}}}{2}\ ,
\ee
\bb
\GG_{\hat{a}\hat{b}\hat{c}\hat{d}\hat{e}}|_0=G_{\hat{a}\hat{b}\hat{c}\hat{d}\hat{e}}\ .
\ee
We also define
${\hat{F}}_{\hat{a}\hat{b}\hat{c}}|_0\equiv F_{\hat{a}\hat{b}\hat{c}}$.
And, for completeness, we write the full form of the supervielbein
\bb
E^A_M|_0=\lk(
\begin{array}{ccc}
{\hat{e}}^{\hat{a}}_m & 0 & 0\\
0 & \delta^\alpha_\mu & 0 \\
0 & 0 & -\delta^{\bar{\alpha}}_{\bar{\mu}}
\end{array}
\re)\ ;
\ee
with the zeroth components of the connection
\bb
\Omega_{cA}^B|_0=
\omega_{c,A}^B
+\mbox{U(1) connection}\ ;
\ee
\bb
\Omega_{\alpha,A}^B|_0=
\Omega_{\bar{\alpha},A}^B|_0=0\ ;
\ee
and the other combinations of indices being zero.

In addition, we will need the zeroth components of the
Riemann and torsion superfields, as well as various spinorial components
of all the superfields. To make things even worse, 
various first and second order
spinorial derivatives of the superfields will also be needed; 
\ie\ some of the higher order
terms in the superfield expansions appear in the sigma model. 
These can be systematically, albeit sometimes tediously,  obtained
by juggling the superspace 
Bianchi identities.
We will present the relevant  
pieces as we need them, instead of cataloging an incomplete set of 
lengthy equations out of context.

\subsection{The IIB string worldsheet in superspace}

The action of the IIB string in a background represented
by the superfields listed above was written in~\cite{GHMNT}
\bb\label{susyaction}
I=\int d^2\sigma\ \lk\{
\frac{1}{2} \sqrt{-h} h^{ij} \Phi V_i^{\hat{a}} V_j^{\hat{b}} \eta_{\hat{a}\hat{b}}
+\frac{1}{2} \varepsilon^{ij} V_i^B V_j^A \BB_{AB}
\re\}\ ,
\ee
with\footnote{Note that the index $\hat{a}$ here runs over all ten spacetime directions including the light-cone.}
\bb
V_i^A\equiv \del_i z^M E_M^A=
\lk\{V_i^{\hat{a}},V_i^\alpha,V_i^{\bar{\alpha}}
\re\}\ ,
\ee
and
\bb
d\BB=\hat{\FF}+\bar{{\hat{\FF}}}\ ,
\ee
\bb
\Phi=\omega=\bar{\Phi}\ .
\ee
The last statement is needed to assure that the action
is $\kappa$ symmetric.
The task is to expand this action in component form. This is generally
a messy matter, which, however, can be achieved using the
algorithm of normal coordinate expansion.

\subsection{The method of normal coordinate expansion in superspace}

Normal coordinate expansion, as applied to bosonic sigma models,
was first developed in~\cite{MUKHINLSM}. In these scenarios, the method helped
to unravel some of
the dynamics of highly non-linear theories approximately, 
as expanded near a chosen point on the target manifold.
In the superspace incarnation, the technique is most
powerful when used to expand 
an action only in a submanifold of the target superspace. In particular,
expanding in the fermionic variables only, with the space coordinate 
left arbitrarily, the expansion truncates by virtue of the Grassmanian
nature of the fermionic coordinates; leading to an exact expression
for the action in component form. This can also be applied of course to
the action or equations of motion
for the background superfields as well, and the technique 
has been demonstrated in this context in many examples. As for the 
IIB sigma model, the expansion has been applied in~\cite{GMPRV}, to 
expand however
the action in all of superspace, leading to a linearized approximate
form that can be used to study quantum effects. Our interest is to
get to an exact expression for~\pref{susyaction} in component form, by fixing
the $\kappa$ symmetry and leaving the space coordinates arbitrary. 
This approach was applied to the Heterotic string in~\cite{ATICKDHAR}. 
There, the absence of RR fields made the discussion considerably 
simpler. Our approach will probe in this respect a new class of couplings
by the use of this method. However, many simplifications and techniques
we will use are direct generalizations of the corresponding 
methods applied in~\cite{ATICKDHAR}. 
First, we briefly review the normal coordinate
expansion method in superspace. The reader is referred 
to~\cite{GRISARU,ATICKDHAR}
for more information.

The superspace coordinates are written as
\bb
Z^M=Z_0^M+y^M\ .
\ee
We choose
\bb
Z_0^M=\lk(x^m,0\re)\ ,\ \ \ 
y^M=\lk(0,y^\mu\re)\ ,
\ee
hence expanding only in the fermionic submanifold.
The action is then given by 
\bb\label{edelta}
I[Z]=e^{\Delta} I[Z_0]\ ,
\ee
with the operator $\Delta$ defined by
\bb
\Delta\equiv\int d^2 \sigma\ y^A(\sigma) {\hat{D}}_A(\sigma)\ ,
\ee
and ${\hat{D}}_A$ being the supercovariant derivative. This derivative
is notationally distinguished from $D_A$ appearing elsewhere in this work in that
it involves the standard connection {\em and} the U(1) connection.
And we use the supervielbein to translate between tangent space
and superspacetime indices
\bb
{\hat{D}}_A\equiv E^N_A(Z_0) {\hat{D}}_N\ ,\ \ \ 
y^N\equiv y^A E_A^N\ .
\ee
For our choice of expansion variables, we then have
\bb
y^{\hat{a}}=0\ ,\ \ \ 
y^\alpha=y^\mu\delta_\mu^\alpha\equiv \theta^\alpha\ ,\ \ \ 
y^{\bar{\alpha}}=y^{\bar{\mu}}\delta_{\bar{\mu}}^{\bar{\alpha}}\equiv 
\theta^{\bar{\alpha}}\ .
\ee

The power of this technique is that it renders the process of expansion
{\em algorithmic}. A set of rules
can be taught say to any well-trained mammal; 
in principle, human intervention
(for that matter the same mammal may be used again) is needed only
at the final stage when
Bianchi identities may be used to determine some of
the expansion terms. The rules are as follows:

\begin{itemize}
\item Due to the definition of the normal coordinates, we have
\bb\label{rule1}
\Delta y^A=0\ .
\ee

\item Using super-Lie derivatives, it is straightforward to derive
\bb
\Delta V_i^A={\hat{D}}_i y^A+V_i^C y^B T_{BC}^A\ .
\ee

\item And the following identity is needed beyond second order
\bb
\Delta\lk({\hat{D}}_iy^A\re)=y^B V_i^D y^C R_{CDB}^{A}\ .
\ee

\item Finally, when we apply $\Delta$ to an arbitrary tensor with
tangent space indices, we get simply
\bb\label{rule4}
\Delta X_{BC..}^{DE..}=y^A D_A X_{BC..}^{DE..}\ .
\ee
\end{itemize}
In the next section, we outline the process of applying these rules 
to~\pref{susyaction}.

\section{Unraveling the action}

There are three sets of difficulties that arise when attempting to
apply the normal coordinate expansion to~\pref{susyaction}. First, a priori, 
we need to expand to order $2^{5}$ in $\theta$ before the expansion
truncates. This problem 
is remedied simply by fixing the $\kappa$ symmetry
with the light-cone gauge, truncating the action to quartic
order in $\theta$, as we will show below.
The second problem is that the expansion terms will need first and second
order fermionic derivatives of the superfields. This requires us to play around
with some of the Bianchi identities to extract the additional information.
The process is somewhat tedious, but straightforward.
The third problem is computational. Despite the simplifications induced
by the light-cone gauge choice, and the algorithmic nature of the process,
it turns out that the task is virtually impossible to perform by a human hand,
while still maintaining some level of
confidence in the result. On average $10^4$
terms arise at various stages of the computation. The use of the computer
for these analytical manipulations greatly simplifies the problem. However,
we find that, even with this help, the complexity is such that computing
time may be of order of many months, unless the task is approached with a set
of somewhat smarter computational steps and unless one makes use of
the simplifications that arise from the
conditions imposed on the background fields as listed in the Introduction.
We do not present all the messy
details of these nuances, concentrating instead on the general protocol.

At zeroth order, the action is simply
\bb
I^{(0)}=I|_0=\int d^2 \sigma \lk\{
\frac{1}{2} \sqrt{-h} h^{ij}\omega V_i^{\hat{a}} V_{j\hat{a}}
+\frac{1}{2} \varepsilon^{ij}  V_i^{\hat{b}} V_j^{\hat{a}} b_{\hat{a}\hat{b}}^{(1)}
\re\}\ .
\ee
Note that this is written with respect to the Einstein frame
metric.

At first order in $\Delta$, the action becomes
\bbb
I^{(1)}&=&\Delta I=
\int\ d^2\sigma \lk\{
\frac{1}{2} \sqrt{-h} h^{ij} (\Delta\Phi) V_i^{\hat{a}} V_j^{\hat{b}} \eta_{\hat{a}\hat{b}}
+\sqrt{-h} h^{ij} \Phi (\Delta V_i^{\hat{a}}) V_j^{\hat{b}} \eta_{\hat{a}\hat{b}}\re.\nonumber \\
&+&\lk.\frac{1}{2} \varepsilon^{ij} V_i^B V_j^A y^C \HH_{CAB}
\re\}\label{deltaI}\ ,
\eee
with
\bb
\HH\equiv d\BB\ .
\ee
This result is not evaluated at for $\theta\rightarrow 0$ yet as
further powers of $\Delta$ will hit it.

\subsection{Fixing the $\kappa$ symmetry}

Matters are simplified if we analyze the form of the action we
expect from this expansion once the $\kappa$ symmetry is fixed.
This will help us avoid manipulating many of the terms that will
turn out to be zero in the light-cone gauge anyways. To fix the $\kappa$
symmetry, we define
\bb\label{sigmapm}
\sigma^{\pm}\equiv \frac{1}{2} \lk(\sigma^0\pm \sigma^{\hat{a}} \re)\ ,
\ee
where $\hat{a}$ is some chosen direction in space.
For conventions on spinors, the reader is referred to Appendix A
and~\cite{HOWEWEST}. We choose the spacetime fermions to satisfy the condition
\footnote{
Alternatively, we can choose~\cite{PESANDO}
\bb
\sigma^{\pm}\equiv \frac{1}{2} \lk(\sigma^{\hat{a}}\pm i \sigma^{\hat{b}}\re)\ ,
\ee
with $\hat{a}$ and $\hat{b}$ being two arbitrary space directions. We can then impose
\bb
\sigma^+\theta=\sigma^-\bar{\theta}=0\ .
\ee
It can be seen that this choice leads to a more complicated
expansion for the action. It may still be necessary to consider 
such choices for other classes of background fields than those
we focus on in this work.
}
\bb\label{LC}
\sigma^+\theta=\sigma^+\bar{\theta}=0\ .
\ee

Consider first all even powers of $\theta$. These will necessarily 
come in one of the following bilinear combinations
\bb\label{bi1}
A^{ab}\equiv \theta \sigma^{-ab}\theta\ ,\ \ \ 
\bar{A}^{ab}=\bar{\theta}\sigma^{-ab}\bar{\theta}\ ;
\ee
\bb\label{bi2}
B\equiv\bar{\theta} \sigma^- \theta\ ,\ \ \ 
B^{ab}\equiv\bar{\theta}\sigma^{-ab}\theta\ ,\ \ \ 
B^{abcd}\equiv \bar{\theta}\sigma^{-abcd}\theta\ .
\ee
In these expressions, condition~\pref{LC} has been used, and 
the Latin indices $a,b,c,d$ are transverse to the light cone
directions. Furthermore, because of the self duality condition
\bb
\tilde{\sigma}^{(5)}=\sigma^{(5)}
\ee
we have $B^{abcd}=0$.

Given the symmetry properties of the gamma matrices (see Appendix A),
we also know
\bb
\bar{A}^{ab}=-A^{ab}\ ,\ \ \ \bar{B}=B\ ,\ \ \ 
\bar{B}^{ab}=-B^{ab}\ ,\ \ \ 
\bar{B}^{abcd}=B^{abcd}\ .
\ee

\subsection{The expected form of the action}

First, we note that, given that all background fermions ($\lambda$ and
$\psi_m$) are zero, only even powers of $\theta$ can appear in the
expansion. 
We assume that all background fields have only non-zero
components that are either transverse to the light-cone directions, 
or that the light-cone indices in them
come in pairs; and that all the fields depend
only on the transverse coordinates.  For example, denoting the light-cone
directions by $'+'$ and $'-'$, and all transverse coordinates schematically
by $r$,
all fields can only depend on $r$; and a tensor $X_{abc..}$
can be non-zero only if either all $a,b,c,..$ are transverse; or if
$'+'$ and $'-'$ come as in $X_{-+bc..}$ with $b,c..$ transverse or other
light-cone pairs. 
These conditions lead to a dramatic simplification of the expansion.
In particular, given that a $'-'$ index is to appear in all even powers of
fermion bilinears, as in~\pref{bi1} and~\pref{bi2}, we must pair each bilinear
with a $V_i^a$ to absorb the light-cone index $'-'$. 

Let $\Theta$ represent either $\theta$ or $\bar{\theta}$. For example,
schematically
$\Theta^2\sim \theta^2, \bar{\theta} \theta, \bar{\theta}^2$.
The action consists then of 
terms of form 
$\Theta^{2n} V_i^a V_j^b$, 
$(D\Theta) \Theta^{2n-1} V_i^a$ and
$(D\Theta) (D\Theta) \Theta^{2n}$.
From the expansion algorithm outlined above, with the use of equations
~\pref{rule1}-\pref{rule4}, it is easy to see that
\bbb
\mbox{number of V's}+\mbox{number of }D\Theta\mbox{'s}=2\nonumber
\eee
in each term.
Let's then look at each class of terms separately:
\begin{itemize}
\item For terms of the form
$\Theta^{2n} V^a V^b$, the 
only non-zero combinations are $\Theta^2 V_i^+ V_j^a$ 
and $\Theta^4 V_i^+ V_j^+$.
This means in particular
that the Wess-Zumino term involving $\HH$ in~\pref{deltaI} does not 
contribute at quartic order since we must contract $V_i^+ V_j^+$ by $\sqrt{-h} h^{ij}$. 

\item Terms of the form
$(D\Theta) \Theta^{2n-1} V_i^a$ are zero unless $n=1$, because, otherwise,
there is shortage of $V$s
to absorb all light-cone indices.

\item Terms of the form
$(D\Theta) (D\Theta) \Theta^{2n}$ are zero for all $n$ for the same
reason as above.
\end{itemize}

Hence, the action must have the form
\bb
I\sim \Theta D\Theta+\Theta^2+\Theta^4 V^+ V^+\ ,
\ee
with the quartic piece receiving contributions only from the first
two terms of~\pref{deltaI}.
Hence, the action truncates at quartic order in the fermions. And we focus on expanding only the relevant parts.

\section{More details}

\subsection{The quadratic terms and comparison to literature}

As we expand~\pref{edelta}, the quadratic terms in $\theta$ are very simple 
to handle, and can be done by hand. On finds that zeroth components
of $D\omega$ and $D^2 \omega$ are needed. For these, we note the relation
\bb
d\omega=-\frac{\omega}{2} \lk(P+\bar{P}\re)\ .
\ee
Using the results of~\cite{HOWEWEST}, we get
\bb
{\hat{D}}_\alpha \omega|_0={\hat{D}}_{\bar{\alpha}}\omega |_0=0\ .
\ee
\bb
{\hat{D}}_{\alpha} {\hat{D}}_{\beta} \omega|_0=-\omega 
\frac{i}{24} \sigma^{\hat{a}\hat{b}\hat{c}}_{\alpha\beta} F_{\hat{a}\hat{b}\hat{c}}\ ,\ \ \ 
{\hat{D}}_{\bar{\alpha}} {\hat{D}}_{\bar{\beta}} \omega|_0=-\omega 
\frac{i}{24} \sigma^{\hat{a}\hat{b}\hat{c}}_{\alpha\beta} \bar{F}_{\hat{a}\hat{b}\hat{c}}\ ,
\ee
\bb
{\hat{D}}_{\bar{\alpha}} {\hat{D}}_{\beta} \omega|_0=-\omega 
\frac{i}{2} \sigma^{\hat{a}}_{\alpha\beta} P_{\hat{a}}|_0\ ,\ \ \ 
{\hat{D}}_{{\alpha}} {\hat{D}}_{\bar{\beta}} \omega|_0=-\omega 
\frac{i}{2} \sigma^{\hat{a}}_{\alpha\beta} \bar{P}_{\hat{a}}|_0\ .
\ee
Note that the supercovariant derivative $\hat{D}_A$ is associated
with the standard supergravity superconnection {\em plus} the
$U(1)$ contribution, as discussed in~\cite{HOWEWEST}.
In these equations, a Latin indices $\hat{a},\hat{b},\hat{c},\cdots$ run over all ten spacetime directions,
the transverse and the light-cone.
In the Wess-Zumino term, we need
$D_\alpha \HH_{\beta \hat{a}\hat{b}}|_0$ and
$D_{\bar{\alpha}} \HH_{\beta \hat{a}\hat{b}}|_0$.
These are found
\bb
{\hat{D}}_\alpha \HH_{\beta \hat{a}\hat{b}}|_0=i\frac{\omega}{2}
\sigma_{\hat{a}\hat{b}\beta}^{\hspace{15pt}\gamma}
\sigma^{\hat{c}}_{\alpha\gamma} \bar{P}_{\hat{c}}|_0
\ee
\bb
{\hat{D}}_{\bar{\alpha}} \HH_{\beta \hat{a}\hat{b}}|_0=i\frac{\omega}{24}
\sigma_{\hat{a}\hat{b}\beta}^{\hspace{15pt}\gamma} 
\sigma^{\hat{c}\hat{d}\hat{e}}_{\bar{\alpha}\gamma} \bar{F}_{\hat{c}\hat{d}\hat{e}}\ .
\ee

Putting things together, we get a kinetic part for the fermions of the form
\bb
-\frac{i}{2} \omega V_{\hat{a}\,i} \Theta^{ij} \sigma^{\hat{a}} {\hat{D}}_j \overbar{\theta}
+\mbox{c.c.}=
-\frac{i}{2} \omega V_{\hat{a}\,i} \Theta^{ij} \sigma^{\hat{a}} D_j \overbar{\theta} 
-\frac{1}{2} \omega Q_{\hat{b}} V_{\hat{a}\,i}\,V^{\hat{b}}_j \Theta^{ij}\sigma^{\hat{a}} \overbar{\theta}+\mbox{c.c.}\ ,
\ee
where the second term arises from the U(1) connection (the $\theta$'s are charged under this U(1)~\cite{HOWEWEST}), and we have defined
\bb
\Theta^{ij}\equiv \sqrt{-h}\,h^{ij} \theta-\varepsilon^{ij} \overbar{\theta}\ .
\ee
To compare with the literature, we want to write the quadratic part in the string frame.
Using
\bb
D_i\theta={\tilde{D}}_i \theta+\frac{1}{8} V_{i\,\hat{b}} \lk(P_{\hat{a}}+\overbar{P}_{\hat{a}}\re)\sigma^{\hat{a}\hat{b}} \theta
\ee
we can write things in terms of the string frame covariant derivative $\tilde{D}$ with metric
\bb
G^{(str)}_{mn}=\omega\, g_{mn}
\ee
We note in particular the relation $\del_{\hat{a}}\ln\omega=-(1/2) (P_{\hat{a}}+\overbar{P}_{\hat{a}})$.
Switching to the string frame rescales the vielbein and hence the
various fields in the action as well
\bb
V_i^{\hat{a}}\rightarrow \omega^{-1/2}V_i^{\hat{a}}\ \ \ ,\ \ \ P\rightarrow \omega^{1/2} P\ \ \ ,\ \ \ F\rightarrow \omega^{3/2} F\ \ \ ,\ \ \ G\rightarrow \omega^{5/2} F\ .
\ee
Finally, we rescale the spinors $\theta\rightarrow \omega^{-1/4} \theta$ so as to
canonically normalize the kinetic term. Collecting all this together, and using
the properties of our gamma matrices, we write the action as
\bb
\SS_{quad}=\int d^2\sigma\,\lk(\II_{D\theta}+\II_F+\II_G\re)+\mbox{c.c.}
\ee
with
\bbb\label{err}
\II_{D\theta}&=&-\frac{i}{2} V_{\hat{a}\,i} \Theta^{ij} \sigma^{\hat{a}} \tilde{D}_j \overbar{\theta}\nonumber \\
&-&\frac{1}{2}V_{\hat{a}\,i} V^{\hat{b}}_j Q_{\hat{b}} \Theta^{ij}\sigma^{\hat{a}}\overbar{\theta}
+\frac{1}{4} V_{\hat{a}\,i} V_{\hat{b}\,j} Q_{\hat{c}} \Theta^{ij}\sigma^{\hat{a}\hat{b}\hat{c}}\overbar{\theta}
+\frac{1}{4} V_{\hat{b}\,i} V^{\hat{b}}_j Q_{\hat{a}} \Theta^{ij}\sigma^{\hat{a}} \overbar{\theta}
\eee
and
\bbb
\II_F&=&i \frac{\omega}{32}\, V_{\hat{a}\,j} V_{\hat{d}\,i} \Theta^{ij}\sigma^{\hat{b}\hat{c}\hat{d}}\theta
\lk(F_{\hat{a}\hat{b}\hat{c}}+3 \overbar{F}_{\hat{a}\hat{b}\hat{c}}\re)
-i \frac{\omega}{32}\, V_{\hat{a}\,i} V_{\hat{d}\,j} \Theta^{ij}\sigma^{\hat{b}\hat{c}\hat{d}}\theta
\lk(F_{\hat{a}\hat{b}\hat{c}}-\overbar{F}_{\hat{a}\hat{b}\hat{c}}\re)\nonumber \\
&+&i \frac{\omega}{96}\, V_{\hat{d}\,i} V^{\hat{d}}_{j} \Theta^{ij}\sigma^{\hat{a}\hat{b}\hat{c}}\theta
\lk(F_{\hat{a}\hat{b}\hat{c}}-\overbar{F}_{\hat{a}\hat{b}\hat{c}}\re)+
i \frac{\omega}{8}\, V^{\hat{a}}_{i} V^{\hat{b}}_{j} \Theta^{ij}\sigma^{\hat{c}}\theta
F_{\hat{a}\hat{b}\hat{c}}
\eee
\bbb
\II_G&=&-\frac{\omega}{96} V^{\hat{a}}_{i} V^{\hat{b}}_{j} \Theta^{ij}\sigma^{\hat{c}\hat{d}\hat{e}}\overbar{\theta}
G_{\hat{a}\hat{b}\hat{c}\hat{d}\hat{e}}\ .
\eee
These expressions agree with~\cite{OTHER} except for a numerical factor in one of the terms. In~\cite{OTHER}, the first term of the second line of~\pref{err} appears
with an additional factor of 2. This term arises from the U(1) charge associated with the spinor. 
We believe that the discrepancy is accounted for by a typo in~\cite{OTHER} 
(perhaps related to adding the complex conjugate piece to the action). Otherwise, our
expressions are identical. 
We conclude that the result, to quadratic order in the spinors, agrees with the literature\footnote{The minor issue regarding the coefficient of the U(1) charge cannot
be settled through comparison to other sources of literature because this term vanishes for cases involving AdS backgrounds.}.

\subsection{The quartic terms}

At quartic order in $\theta$, the action is much more difficult 
to find. Indeed, the use of computation by machine becomes
necessary. We do not present all the details, but only
some of the important relations that are needed to check the results. In this section,
to avoid clutter in index notation, indices $a,b,c,\cdots$ will run over all ten spacetime
directions as opposed to using $\hat{a},\hat{b},\hat{c},\cdots$ as we did in the rest of the
paper.

First derivatives of some of the Riemann tensor components arise; 
particularly, 
${\hat{D}}_{\bar{\alpha}} {\hat{R}}_{\beta a\gamma_1}^{\gamma_2}$ and
${\hat{D}}_{\alpha} {\hat{R}}_{\bar{\beta} a\gamma_1}^{\gamma_2}$.
Using the results of~\cite{HOWEWEST}, it is straightforward to find
\bb
{\hat{D}}_{\alpha} {\hat{R}}_{\bar{\beta} a\gamma_1}^{\gamma_2}|_0=
\frac{i}{8} \sigma^{cd\gamma_2}_{\gamma_1} 
\lk(
\sigma_{a\bar{\beta}\delta}{\hat{D}}_{\alpha}T_{cd}^\delta+
\sigma_{c\bar{\beta}\delta}{\hat{D}}_{\alpha}T_{ad}^\delta+
\sigma_{d\bar{\beta}\delta}{\hat{D}}_{\alpha}T_{ca}^\delta
\re)|_0
+\frac{i}{2} \delta_{\gamma_1}^{\gamma_2} P_a \sigma^b_{\alpha\bar{\beta}}
\bar{P}_b|_0\ ;
\ee
\bb
{\hat{D}}_{\bar{\alpha}} {\hat{R}}_{{\beta} a\gamma_1}^{\gamma_2}|_0=
-\frac{i}{8} \sigma^{cd\gamma_2}_{\gamma_1} 
\lk(
\sigma_{a{\beta}\bar{\delta}}{\hat{D}}_{\bar{\alpha}}T_{cd}^{\bar{\delta}}+
\sigma_{c{\beta}\bar{\delta}}{\hat{D}}_{\bar{\alpha}}T_{ad}^{\bar{\delta}}+
\sigma_{d{\beta}\bar{\delta}}{\hat{D}}_{\bar{\alpha}}T_{ca}^{\bar{\delta}}
\re)|_0
-\frac{i}{2} \delta_{\gamma_1}^{\gamma_2} {\bar{P}}_a \sigma^b_{\bar{\alpha}{\beta}}
{P}_b|_0\ .
\ee
We note the distinction between $R$ and $\hat{R}$; the latter includes
the curvature from the $U(1)$ gauge field, as defined in~\cite{HOWEWEST}.
To avert confusion, we also note that the covariant derivative ${\hat{D}}_A$ is
with respect to $\hat{R}$; whereas the one appearing elsewhere in the text
as $D$ does not involve the $U(1)$ connection. This aspect of our notation then differs slightly 
from that of~\cite{HOWEWEST}.

We need a series of first spinorial derivatives of the torsion.
For these, we need to use the Bianchi identity
\bb
\sum_{(ABC)} {\hat{D}}_A T_{BC}^D
+T_{AB}^E T_{EC}^D
-{\hat{R}}_{ABC}^D=0\ ,
\ee
where the sum is over graded cyclic permutations.
We then find 
\bb
{\hat{D}}_\alpha T_{cd}^\delta|_0=
R_{cd\alpha}^\delta
-{\hat{D}}_dT_{\alpha c}^\delta-{\hat{D}}_cT_{d\alpha}^\delta
+2 T_{\alpha [d}^{\bar{\beta}} T_{c]\bar{\beta}}^\delta
-2 T_{\alpha [d}^{{\beta}} T_{c]{\beta}}^\delta
+\delta_{\alpha}^{\delta} {\bar{P}}_{[c} P_{d]}\ ,
\ee
and
\bb
{\hat{D}}_{\bar{\alpha}} T_{bc}^{\bar{\delta}}|_0=
-{\hat{D}}_bT_{c\bar{\alpha}}^{\bar{\delta}}
-{\hat{D}}_cT_{\bar{\alpha}b}^{\bar{\delta}}
+R_{bc\bar{\alpha}}^{\bar{\delta}}
+2 T_{\bar{\alpha}[c}^{\bar{\gamma}}T_{b]\bar{\gamma}}^{\bar{\delta}}
-2 T_{\bar{\alpha}[c}^{{\gamma}}T_{b]{\gamma}}^{\bar{\delta}}
+\delta_{\bar{\alpha}}^{\bar{\delta}} {\bar{P}}_{[b} P_{c]}\ .
\ee
We also have
\bb
{\hat{D}}_{\alpha} T_{\bar{\beta} \bar{\gamma}}^{\delta}|_0=
-\frac{i}{24} \sigma^d_{\bar{\beta}\bar{\gamma}}
\sigma_{d}^{\delta\beta}
\sigma^{abc}_{\alpha\beta} F_{abc}
+\frac{i}{24} \delta^{\delta}_{\bar{\beta}}
\sigma^{abc}_{\alpha\bar{\gamma}} F_{abc}
+\frac{i}{24} \delta^{\delta}_{\bar{\gamma}}
\sigma^{abc}_{\alpha\bar{\beta}} F_{abc}\ .
\ee
In all these and subsequent equations, the right hand sides are to be
evaluated as zeroth order in $\theta$.

As if first derivatives are not enough of a mess, two derivatives
of the torsion are also needed. For example,
${\hat{D}}_\alpha {\hat{D}}_\beta T_{\bar{\gamma},a}^{\delta}$ arises and is found
\bbb
& &{\hat{D}}_\alpha {\hat{D}}_\beta T_{\bar{\gamma} a}^{\delta}|_0=
-\frac{3}{16}\sigma_{\bar{\gamma}}^{de\delta}\lk(
-\frac{1}{32} 
\KK_{ade\beta}^{\hspace{20pt}\gamma} {\hat{D}}_{\alpha}{\hat{D}}_\gamma\omega
+3 P_{[a}\sigma_{de]\beta}^{\hspace{18pt}\gamma} {\hat{D}}_\alpha {\hat{D}}_\gamma\omega
+3 i \sigma_{[a\beta\gamma}{\hat{D}}_\alpha T_{de]}^\gamma
\re)\nonumber\\
&-&\frac{1}{48} \sigma_{a\bar{\gamma}}^{\hspace{8pt}cde\delta}\lk(
-\frac{1}{32}
\KK_{cde\beta}^{\hspace{20pt}\gamma} {\hat{D}}_{\alpha}{\hat{D}}_\gamma\omega
+3 P_{[c}\sigma_{de]\beta}^{\hspace{18pt}\gamma} {\hat{D}}_\alpha {\hat{D}}_\gamma\omega
+3 i \sigma_{[c\beta\gamma}{\hat{D}}_\alpha T_{de]}^\gamma
\re)\ ,
\eee
where we define the matrix
\bb
\KK_{cde}\equiv \sigma_{cdefgh} \bar{F}^{fgh}
+3 \bar{F}_{[c}^{\hspace{7pt}fg}\sigma_{de]fg}
+52 \bar{F}_{[cd}^{\hspace{12pt}f} \sigma_{e]f}
+28 \bar{F}_{cde}\ .
\ee
To find
${\hat{D}}_{\bar{\alpha}} {\hat{D}}_\beta T_{\gamma_1 a}^{\gamma_2}$,
we use the standard statement
\bb
[{\hat{D}}_A,{\hat{D}}_B\}=-T_{AB}^C {\hat{D}}_C
-{\hat{R}}_{ABC}^D\ .
\ee
And we get
\bb
{\hat{D}}_{\bar{\alpha}} {\hat{D}}_\beta T_{\gamma_1 a}^{\gamma_2}|_0=
-T_{\bar{\alpha}\beta}^b {\hat{D}}_b T_{\gamma_1 a}^{\gamma_2}
+R_{\bar{\alpha}\beta\gamma_1}^\delta T_{\delta a}^{\gamma_2}
+R_{\bar{\alpha}\beta a}^b T_{\gamma_1 b}^{\gamma_2}
-T_{\gamma_1 a}^{\delta} R_{\bar{\alpha}\beta\delta}^{\gamma_2}
-{\hat{D}}_\beta {\hat{D}}_{\bar{\alpha}} T_{\gamma_1 a}^{\gamma_2}\ .
\ee
We need
${\hat{D}}_{\alpha} {\hat{D}}_{\bar{\beta}} T_{\gamma_1 a}^{\gamma_2}$, which is
\bbb
{\hat{D}}_{\alpha} {\hat{D}}_{\bar{\beta}} T_{\gamma_1 a}^{\gamma_2}|_0&=&
-{\hat{D}}_\alpha {\hat{D}}_{\gamma_1} T_{\bar{\beta} a}^{\gamma_2}
-T_{\bar{\beta}\gamma_1}^b {\hat{D}}_{\alpha} T_{ba}^{\gamma_2}
-{\hat{D}}_{\alpha} R_{\bar{\beta} a \gamma_1}^{\gamma_2} \nonumber \\
&-&T_{\gamma_1 a}^{\bar{\alpha}} \sigma^{b}_{\bar{\alpha}\bar{\beta}}
\sigma_{b}^{\gamma_2 \delta} {\hat{D}}_{\alpha} {\hat{D}}_{\delta} \omega
+T_{\gamma_1 a}^{\bar{\alpha}} \delta^{\gamma_2}_{\bar{\alpha}}
\delta_{\bar{\beta}}^\delta {\hat{D}}_{\alpha} {\hat{D}}_\delta \omega
+T_{\gamma_1 a}^{\bar{\alpha}} \delta_{\bar{\beta}}^{\gamma_2}
\delta_{\bar{\alpha}}^\delta {\hat{D}}_\alpha {\hat{D}}_\delta \omega\ .
\eee

Finally, we collect the zeroth order components of some of the superfields
that arise in the computation as well. These can be found in~\cite{HOWEWEST},
but we list them for completeness:
\bb
T_{\alpha\bar{\beta}}^a|_0=-i\sigma^a_{\alpha\beta}\ .
\ee
\bb
T_{a\beta}^{\bar{\gamma}}|_0=
-\frac{3}{16} \sigma_\beta^{bc\gamma}\bar{F}_{abc}
-\frac{1}{48} \sigma_{abcd\beta}^{\hspace{25pt}\gamma} \bar{F}^{bcd}\ .
\ee
\bb
T_{a\beta}^\gamma|_0=i \sigma^{bcde\gamma}_\beta Z_{abcde}\ .
\ee
\bb
R_{\alpha\beta,ab}|_0=i\frac{3}{4} \sigma^c_{\alpha\beta}
\bar{F}_{abc}
+\frac{i}{24} \sigma_{abcde\alpha\beta}\bar{F}^{cde}\ .
\ee
\bb
R_{\alpha\bar{\beta},ab}|_0=-\frac{1}{24}\sigma^{cde}_{\alpha\beta}
g_{abcde}\ .
\ee
\bb
H_{a\beta\gamma}|_0=-i\omega \sigma_{a\beta\gamma}\ .
\ee
\bb
H_{a\bar{\beta}\bar{\gamma}}|_0=-i\omega \sigma_{a\beta\gamma}\ .
\ee
All other components as they arise in the expansion
are zero.
The final result is given in~\pref{I4}.

\section{Discussion}

In this work, we derived the component form of the IIB worldsheet theory 
in backgrounds involving RR fluxes. In the light-cone gauge, the action was found to truncate to quartic
order in the spacetime spinors. Terms quadratic in the fermions could
be compared to results already existing in the literature; and we concluded
that our computation agrees with the existing results (modulo a term we commented
on in Section 4.1). The quartic interactions terms are most interesting in addressing
issues of integrability of the worldsheet and were computed as well. The complete results
were summarized in 
equations~\pref{I2} and~\pref{I4}. 

The form of our action is such that the spinors
$\theta$ may dynamically acquire a non-trivial vacuum configuration depending
on the strengths of the various background fields.
There is also an interesting coupling to
the covariant derivative of the field strengths $DF$. And it is easy to see that many of the
terms vanish when one considers center of mass motion of the closed
string. An important program
is then to arrange for
simplified semi-classical settings and see how turning on the various couplings
independently affects the vacuum of the worldsheet theory.
This can help us develop intuition 
about the effects of RR fields on closed string dynamics.
We defer such a  complete analysis
to an upcoming work~\cite{VVSNEW}.

Other future directions include writing the IIA action
in a similar manner, or by using T-duality (see, for example,~\cite{CVETIC}). 
Furthermore,
given the algebraic complexity of the computations involved in deriving 
some parts of our action, it can be
useful to have some of the details of our results
checked independently, preferably with different methods. 
Finally, 
it would be helpful to develop general computational techniques that allow us
to analyze, at least semi-classically, dynamics of closed strings
in arbitrary backgrounds - with the RR fields
taken into account. In this regard,
approximation methods such as expansion about center of mass motion
-- which is in some respects an extension of the normal coordinate
expansion technique we used in superspace -- may be used. We hope to address
some of these issues in the future.

\section{Appendix A: Spinors and conventions}

In this appendix, the indices $a,b,c,\cdots$ run over all ten spacetime
directions.
Our spinors are Weyl but not Majorana. They are then complex and have
sixteen components. The associated $16\times 16$ gamma matrices satisfy
\bb
\lk\{\sigma^a,\sigma^b\re\}=2\eta^{ab}\ ,
\ee
with the metric
\bb
\eta_{ab}=diag(+1,-1,-1,...,-1)\ .
\ee
Note that the signature is different from
the standard one in use in modern literature. This is so that
we conform to the equations appearing in~\cite{HOWEWEST}.
Also, the worldsheet metric $h^{ij}$ has signature $(-,+)$
for space and time, respectively.
Throughout,
the reader may refer to~\cite{HOWEWEST} to determine more about the
spinorial algebra and identities that we are using. However, 
we make no distinction between $\sigma$ and $\hat{\sigma}$ as defined
in~\cite{HOWEWEST} as this will be obvious from the context.

We note that
$\sigma^a$, $\sigma^{abcd}$ and $\sigma^{abcde}$ are symmetric; while
$\sigma^{ab}$ and $\sigma^{abc}$ are antisymmetric; and
$\sigma^{abcde}$ is self-dual.

With the choice given in~\pref{sigmapm}, we then have
\bb
\sigma^+\sigma^-+\sigma^-\sigma^+=1\ .
\ee
And complex conjugation is defined so that
\bb
\overline{\sigma^a}=\sigma^a\ .
\ee
Conjugation also implies
\bb
\overline{\theta_1 \theta_2}=\bar{\theta}_2 \bar{\theta}_1\ .
\ee
Finally, antisymmetrization is defined as
\bb
\sigma^{ab}\equiv \sigma^{[a} \sigma^{b]}\ ,
\ee
with a conventional $2!$ hidden by the braces. 

Using the completeness relation and the algebra above, we have,
for any matrix $Q_{\alpha\beta}$ with lower indices
\bb
Q_{\alpha\beta}=\frac{1}{16}\lk(
\mbox{Tr} [Q \sigma_a] \sigma^a_{\alpha\beta}
-\frac{1}{3!} \mbox{Tr} [Q\sigma_{abc}] \sigma^{abc}_{\alpha\beta}
+\frac{1}{5!} \mbox{Tr} [Q\sigma_{abcde}] \sigma^{abcde}_{\alpha\beta}
\re)\ .
\ee
This allows us, for example, to rearrange certain combinations such as
\bb
(\bar{\theta}\sigma^{-(r)}\theta)
(\bar{\theta}\sigma^{-(s)}\theta)
=\frac{1}{2} \frac{sgn(r)}{16^2}
(\bar{\theta}\sigma^-\sigma^{bc}\bar{\theta})
(\theta\sigma^-\sigma^{ef}\theta)
\mbox{Tr}
[\sigma^{bc}\sigma^{(s)}\sigma^{ef}\sigma^{(r)}]\ ,
\ee
\bb
sgn(r)\equiv \lk\{ 
\begin{array}{cc}
+1 & \mbox{for  }r=0 \\
-1 & \mbox{for  }r=2 \\
+1 & \mbox{for  }r=4
\end{array}\ ;
\re.
\ee
this identity arises repeatedly in the computations.
Finally, to avert confusion, we also note the summation convention used
\bb
U^A V_A=U^a V_a+U^\alpha V_\alpha-U^{\bar{\alpha}} V_{\bar{\alpha}}\ .
\ee

\section{Acknowledgments}
I thank P. Argyres, T. Becher, and M. Moriconi for discussions. 
I am grateful to the organizers and staff of IPAM for their warm
hospitality. This work was supported in part by a grant from the NSF.

\providecommand{\href}[2]{#2}\begingroup\raggedright\endgroup

\end{document}